# Ripplocation in graphite nanoplatelets during sonication assisted liquid phase exfoliation

A.V. Alaferdov[a], R. Savu[a], M.A. Canesqui[a], Y.V. Kopelevich[b], R.R. da Silva[b], N.N. Rozhkova[c], D.A. Pavlov[d], Yu.V. Usov[d], G.M. de Trindade[e], S.A. Moshkalev[a,*]

[a]*Center for Semiconductor Components and Nanotechnologies, University of Campinas, Campinas, SP, 13083-870, Brazil*

[b]*The "GlebWataghin" Institute of Physics, University of Campinas, Campinas, SP, 13083-859, Brazil*

[c]*Institute of Geology, Karelian Research Centre, Russian Academy of Sciences, Petrozavodsk, Russia*

[d]*Lobachevsky State University of Nizhni Novgorod, Nizhni Novgorod, 603950, Russia*

[e]*Nacional de Grafite Ltda, Itapecerica, MG, 35550-000, Brazil*

**Abstract**

Defects induced by liquid-phase exfoliation of graphite using sonication were studied. It was shown that localized impact by cavitation shock waves can produce bulk ripplocations and various types of dislocations in graphite nanoplatelets. Formation of ripples is more pronounced in large aspect (length/width) ratio platelets or nanobelts. Quasi-periodical ripple systems were observed in many nanobelts after sonication. Mechanism of formation of ripples and dislocations during sonication was proposed. Surprisingly, fast high-temperature processing was found to anneal most of defects. This is consistent with our observations that defects associated with ripplocations are strongly localized and thus can be fast annealed.

* Corresponding author. Tel: +55-19-3521-5213. E-mail: stanisla@unicamp.br (Stanislav Moshkalev)

Graphite exfoliation in a liquid phase is the main approach used currently for mass production of graphene. Among the methods used for liquid-phase graphite exfoliation, ultrasound processing (sonication) in different solvents is the most popular [1–3]. Other methods include dry and wet ball milling as well as a group of emerging methods based on fluid dynamics [1,4] (see also Ref. [1] in Supporting Information). The popularity of sonication is due to its low cost, scalability, versatility and ability to produce single layer, a few layer and multilayer graphene sheets (or, graphite nanoplatelets). The exfoliation process using sonication is based on the cavitation mechanism when microbubbles are formed close to (or on) the surface of graphite flakes. When these bubbles collapse, secondary shock waves and microjets hit the flake surface resulting in extremely fast and strong impact, with local pressures and temperatures as high as tens of MPa and several thousand K, respectively, within a microsecond scale time. The resulting tensile stress produced in the flakes eventually leads to their sequential exfoliation during multiple cavitation events. However, the process is so harsh that it inevitably produces various types of defects in layered graphene sheets, and the density of these defects may depend on the intensity and frequency of the ultrasound source, duration of the process, temperature and nature of solvent, etc. However, studies of the effect of sonication of the exfoliation process are still scarce.

In the present work, we studied structural defects induced by sonication processing in multilayer graphene and graphite nanoplatelets (GNP) produced from natural graphite. For this, we used XRD, Raman, SEM and TEM techniques were also examined. The defects we found in nanoplatelets include various types of dislocations (localized deformations) as well as periodical deformations also known as ripplocations (formation of periodical ripples). Surprisingly, most of defects were found to be successfully healed after short time thermal processing in inert atmosphere at elevated temperatures (up to 2950ºC). Based on these findings, we propose here the mechanism responsible for the formation of defects during sonication.

GNP were received from Nacional de Grafite, Brazil [5] in suspensions in 2-propanol (1 mg /mL) or in a powder form after drying. Analytical grade 2-propanol (IPA) was acquired from Sigma-Aldrich and used as received. Graphite nanoplatelets were produced by liquid phase chemically assisted sonication-induced exfoliation of natural graphite [2] and may be in a form of either graphite nanobelts (GNBs) with high length/width aspect ratio or graphite nanoflakes (GNFs) of irregular shape, both with micrometer lateral sizes and thicknesses ranging from a few to tens of nanometers. Nanobelts have widths (W) and lengths (L) ranging from 1 to 5 µm and from 5 to 50 µm, respectively, so that the aspect ratio L/W is usually close to 10. For comparisons, we also used original (not processed) polycrystalline mm-sized samples of natural graphite.

The powders containing GNPs were submitted to thermal treatment in vacuum or inert (Ar) atmospheres, and the comparative analyses of their properties before/after thermal processing was performed. Higher temperatures (from 1500 to 2950°C) were used for short time treatment in Ar, note that longer treatments were not available with the equipments used. 1500°C treatment was carried out in a Pt/Rh crucible with 10-minute soak (maximum temperature) time. Higher temperatures were applied in a graphite crucible with 10-seconds soak time. Longer thermal treatments were carried out at 700, 950 °C in a high vacuum chamber (~$10^{-6}$ Torr base pressure) equipped with 1-inch diameter heater for 2 hours.

For further processing, samples were first dispersed in IPA using sonication bath at ultrasound power of 100 W and frequency of 37 kHz, with typical process times about 10 hours. For individual GNP characterization, strongly diluted suspensions (concentration of 10 µg/mL) were deposited onto holey amorphous carbon TEM grids (SPI Lacey carbon coated 300 mesh Cu PK/100) or thermally oxidized silicon substrates. The individual GNPs were characterized using scanning electron microscopy (SEM, Nova 200 Nanolab, FEI), high resolution transmission electron microscopy (HR-TEM, JEM-2100F, JEOL). The crystalline quality of individual GNPs and thin films based on GNPs (also known as buckypapers) were analyzed by confocal micro-Raman spectroscopy (NT-MDT NTEGRA Spectra, with a 473 nm laser). The crystallographic parameters of the above materials were determined by powder X-ray diffraction (XRD) method using Bruker D2 PHASER, Cu Kα radiation λ = 0.15418 nm.

The results of comparisons of X-ray diffraction patterns for (002) and (004) peaks for samples before and after thermal processing are shown in Fig.1. Deconvolutions of the spectra (examples are shown in Fig.S1, Supporting Information) were carried out to show that (002) peak spectra can be considered basically as superposition of two main distinct peaks located at 26.58-26.60° corresponding to hexagonal ABAB stacking or 2H-graphite with interlayer distances about 3.35 Å [6,7] and at 26.10-26.36° corresponding to 3.41-3.38 Å interlayer distance and usually being interpreted as turbostratic or t-graphite [7], (see also references [2-4] in Supporting Information), and to a lesser extent, at 26.37-26.57° corresponding to rhombohedral 3R-structures with interlayer distance about 3.38-3.36 Å [6]. As expected, a narrow (002) 2H-graphite peak at 26.59° (interlayer distance of 3.352 Å) with full width at half maximum (FWHM) of 0.05° was observed for the natural graphite, with small contribution (~1%) from the 3R-graphite. It must be noted that together with the main 2H-graphite peak at 26.60°, a narrow (FWHM of 0.05°) peak at 26.70° was usually detected, corresponding also to ABAB-stacked graphite but with slightly reduced interlayer distance (see references [4,5] in Supporting Information). After sonication, the (002) peak splits basically into two broadened peaks. One of them is shifted to lower angles, centered at 26.1-26.2° (3.40-3.41 Å interlayer distance) and has a FWHM of 0.63°, another peak is centered at 26.60° with a FWHM of 0.31- 0.39°. For the sonicated samples, there is also some contribution from a broad peak at higher angles (around 27.5°) corresponding to lower interlayer distances (3.20-3.25 Å) as observed in some other studies, that can be interpreted as progressive intermediates in graphite-to-diamond phase transition [8]. As shown in Ref. [8], this structure can be formed due to breaking of bonds in carbon rings under compression along $c$ direction and subsequent formation of new bonds between neighboring layers. Note that the maximum peak positions, widths and relative intensities of peaks were found to vary considerably between different sonicated samples (for example, compare curves 2 and 3 in Fig.1).

Strikingly, after fast high-temperature annealing the peak positions and their FWHM were found to recover to a greater extent, coming back close to the values characteristic of natural graphite, see Fig.1 where the spectrum for 10 seconds 2950°C processing is shown. Full comparison of FWHM changes for (002) and (004) peaks can be seen in Table S1 (Supporting Information). Practically the same results were obtained for 1500 °C and even for lower temperatures of processing although for longer time. The main difference with the case of natural graphite is that the resulting (002) peak is asymmetric with considerable contribution from lower angles, that can be probably assigned to 3R-graphite. Together with that, the out-of-plane crystalline size $L_c$, calculated from the (002) peak of XRD spectra, also indicates significant improvement of the crystallinity, increasing twice from 21-26 nm after sonication to 45-50 nm after annealing (see Table S2, Supporting Information).

Comparative study of samples before and after annealing using transmission microscopy also revealed dramatic changes in their crystalline structure. Before annealing, TEM images showed relatively large density of defects, mainly various types of prismatic dislocations localized in small cross-section areas: «step»-type defect - a few graphene layers undergo jump (buckling) along c-axis, «y»-type defects - two layers merge into one layer and «x»-type defects - two layers merge into one layer and split again into two layers (see examples in Fig. 2a-d and Fig. S3, Supporting Information). The linear density of these defects was relatively high, about 10 per each 100 nm of a GNP cross section. Another type of defects («grain»-type) with small apparently amorphized areas were also observed, having sizes of a few layers along c-axis and few nm in a basal plane direction and smaller linear density, being ~1 per 100 nm (Fig. S3). Again, after fast annealing the density of defects was found to decrease by almost about one order of magnitude, a typical example of a cross-section is shown in Fig.2d. More specifically, «grain»-type defects were no longer observed, but a few «x», «y» or «step»-types of defects per 100 nm were still detected. This comparison is based on analysis of more than 15 samples by TEM both before and after annealing.

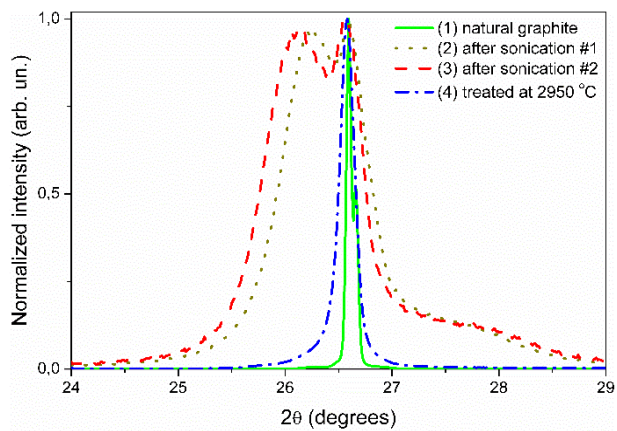

**Figure 1.** X-ray diffraction spectra showing the (002) peak.

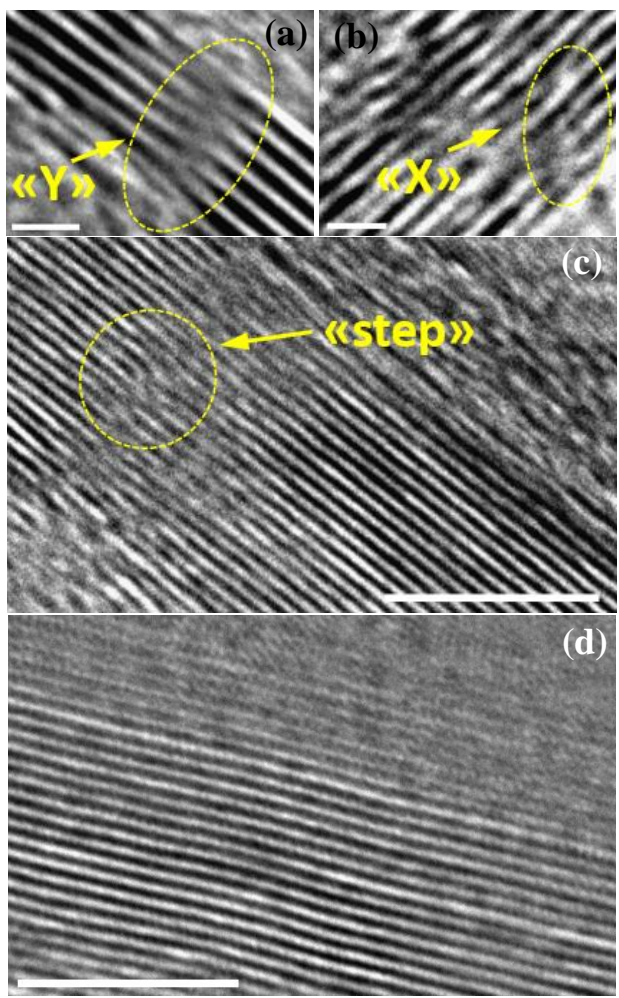

**Figure 2.** Fragments of HR-TEM images of GNP cross sections after sonication (a-c), and after thermal processing (d). Scale bars: 1nm (a) and (b), 5 nm (c) and (d). In (a), (b) and (c) arrows shown type of defects: «Y», «X» and «step» respectively.

Strong improvement of the crystallinity due to annealing was also confirmed by Raman spectroscopy results (see Fig.S4 and Table S3, Supporting Information). Briefly, samples after annealing showed strong reduction of both the D band intensity and FWHM of the G band, confirming the decrease of the defects density [9].

Interestingly, in many samples of nanobelts, formation of quasi-periodical ripples after sonication was observed, with the number of ripples in some nanobelts being as high as 10 (Fig.3). The height of ripples (evaluated from SEM and AFM images) was found to vary considerably, from a few nm up to 100 nm (one example is shown by inset in Fig. 3b). The mean period of such ripples showed strong correlation with the nanobelt width: the smaller the width, the shorter is the period (Fig. 3b). Note that in nanoflakes, characterized by lower aspect (length/width) ratio, ripples are rarely observed, they always have smaller heights (a few nm) and their number in a flake is usually limited by one, located near the central part of the flake (Fig. 4a). In a few cases, formation of several ripples oriented in different directions was observed for larger flakes, as can be seen in Fig. 4b. These findings strongly suggest that formation of ripples is a result of cyclic periodic deformations probably induced by standing waves produced in nanoplatelets by cavitation shock waves. These standing waves might be formed in a platelet due to superposition of waves running in opposite directions along a limited-size elastic sample behaving as a rigid vibrating string. The life time of such waves after the cavitation impact must be quite short (estimated to be in the order of microseconds) as the received energy is transferred rapidly to the surrounding liquid media especially in the direction normal to the sample surface. Large area flakes dissipate the energy of shock-wave induced vibrations to the surrounding liquid faster than narrow nanobelts (with a smaller surface area). Therefore, the vibrations life time must be larger for nanobelts, so it is not surprising that the ripples formed in nanobelts are much higher. It is also possible that these ripples could be an accumulated result of a number of shock waves in sequential cavitation events.

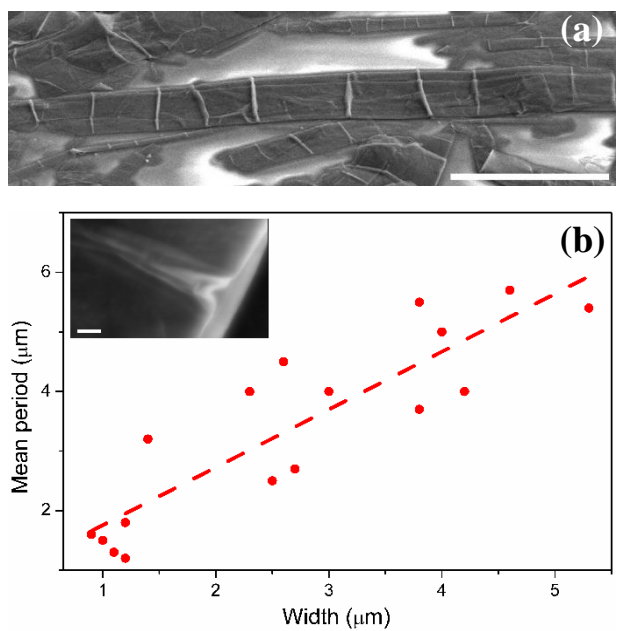

**Fig. 3.** (a) SEM image showing ripplocations in graphite nanobelts (scale bar 10 μm); (b) mean period of ripples vs. the nanoplatelet width, inset – fragment of SEM image with a single ripple profile (scale bar 100 nm).

The ripplocation process in layered materials like graphite under compressive deformation has been analyzed in a recent study [10]. It was considered as a buckling phenomenon that can be developed inside the layered solids (in contrast to more known surface ripplocations) under localized deformations. Starting as small perturbations within a few adjacent layers, these structures tend to combine and grow giving rise to larger kinks or ripples. Formation of such bulk ripplocations with strong out-of-plane deformations can lead to breaking of in-plane bonds that is likely to occur within limited areas (a few layers) extended along c-axis. Further, it is possible that eventual re-bonding between fractured layers can produce structures differing from their initial state, or in other words, produce dislocations like observed in our study. A schematic of such process is shown in Fig. S5, Supporting Information. It is also clear that the spatial extension of such dislocations is determined by the energy deposited in the sample by a shock wave. In the case of highly energetic impact (high effective temperatures, for short time), the sample can be broken into separate parts, otherwise the fractured area is maintained "frozen" inside the sample as can be clearly seen in Figs. 2a-c.

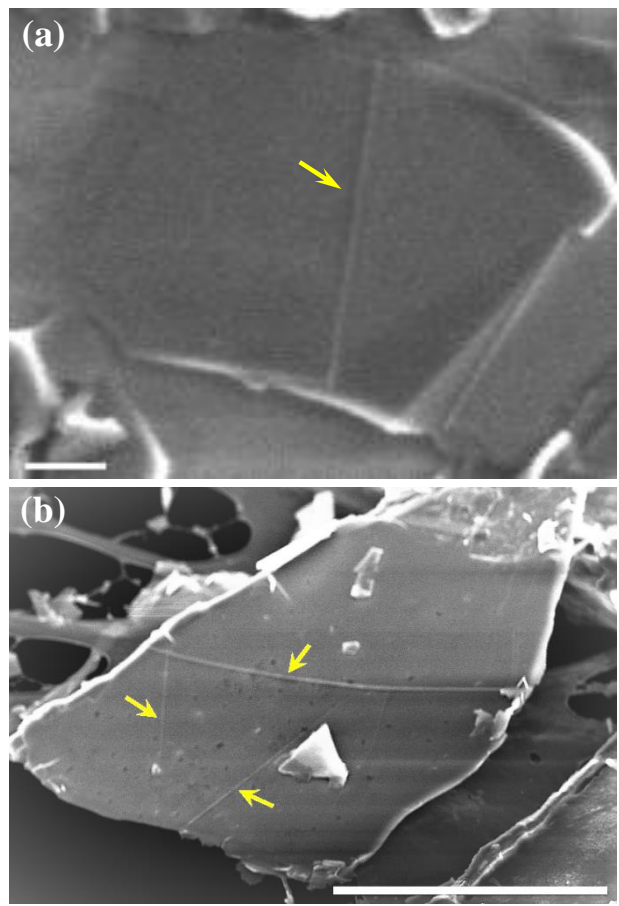

**Fig. 4.** SEM images showed ripples in nanoflakes: one (a) or several (b). Scale bars: 500 nm (a), 10 μm (b). Arrows show positions of ripples.

Further, the fact that the fractured areas are strongly localized, explains the extremely fast annealing of the most of defects under high-temperature processing as evidenced by the results presented in Figs. 1, 2. In this case, relatively slow heating followed by slow cooling of samples (in contrast to the fast heating/fast cooling scenario realized during cavitation events) can create conditions for gradual re-bonding of fractured layers into perfect crystalline structures that are energetically more favorable.

In conclusion, we studied here defects induced by liquid-phase exfoliation of graphite using sonication. It has been shown that due to strong localized impacts produced by cavitation shock waves in graphite nanoplatelets, bulk ripplocations can be developed inside the samples eventually resulting in formation of large-scale ripples, fracture of layers and formation of localized defects like dislocations. Formation of ripples is more pronounced in large aspect (length/width) ratio platelets or nanobelts that behave as vibrating strings after the energetic impact by shock waves. Quasi-periodical ripple systems were observed in many nanobelts indicating formation of standing waves induced by shock waves. The height of ripples in some nanobelt samples reached 100 nm evidencing the extremely high intensity of the ripplocation process and high energy deposited in samples by shock waves. Further, as confirmed by results of XRD, TEM and Raman techniques, fast high-temperature processing successfully anneals most of defects produced by sonication in the nanoplatelets. This is consistent with our observations that the defects associated with ripplocations are strongly localized and thus can be fast annealed. The phenomena observed here for thin graphite samples should be valid also for other layered materials processed in liquid phase for exfoliation.

**Acknowledgements.**

The authors thank CNPq, FAPESP and FINEP (Brazil) for financial support, the CCS Nano staff for technical assistance, and A.V. Bobrov (Lobachevsky State University) for his technical assistance in HR-TEM imaging. This study was also supported by the Ministry of Education and Science of Russian Federation (Project No. 8.1751.2017/PCh).

**Supporting information**

More detailed analysis of XRD spectra, including deconvolution examples of diffractograms; HR-TEM images of GNP cross sections after sonication, Raman spectra analysis, crystalline size calculations from XRD and Raman spectra shown in supporting information. Also we proposed schematics of the buckling process after the shock-wave impact in GNP.

Supporting Information for

**Ripplocation in graphite nanoplatelets during sonication assisted liquid phase exfoliation**

A.V.Alaferdov[a], R.Savu[a], M.A. Canesqui[a], Y.V. Kopelevich[b], R.R. da Silva[b], N.N. Rozhkova[c], D.A. Pavlov[d], Yu.V. Usov[d], G.M. de Trindade[e], S.A.Moshkalev[a]

[a]*Center for Semiconductor Components and Nanotechnologies, University of Campinas, Campinas, SP, 13083-870, Brazil*
[b]*The "GlebWataghin" Institute of Physics, University of Campinas, Campinas, SP, 13083-859, Brazil*
[c]*Institute of Geology, Karelian Research Centre, Russian Academy of Sciences, Petrozavodsk, Russia*
[d]*Lobachevsky State University of Nizhni Novgorod, Nizhni Novgorod, 603950, Russia*
[e]*Nacional de Grafite Ltda, Itapecerica, MG, 35550-000, Brazil*

**Table of contents.**

In the Supporting Information we present more detailed analysis of XRD spectra, including deconvolution examples of diffractograms (Fig.S1); typical HR-TEM images of GNP cross sections after sonication (Fig.S2), possible classification of dislocation defects in graphite (Fig.S3), Raman spectra analysis (Fig.S4), crystalline sizes calculations from XRD and Raman spectra (Tables S1, S2 and S3). Also we proposed schematic of the buckling process after the shock-wave impact in GNP (Fig.S5).

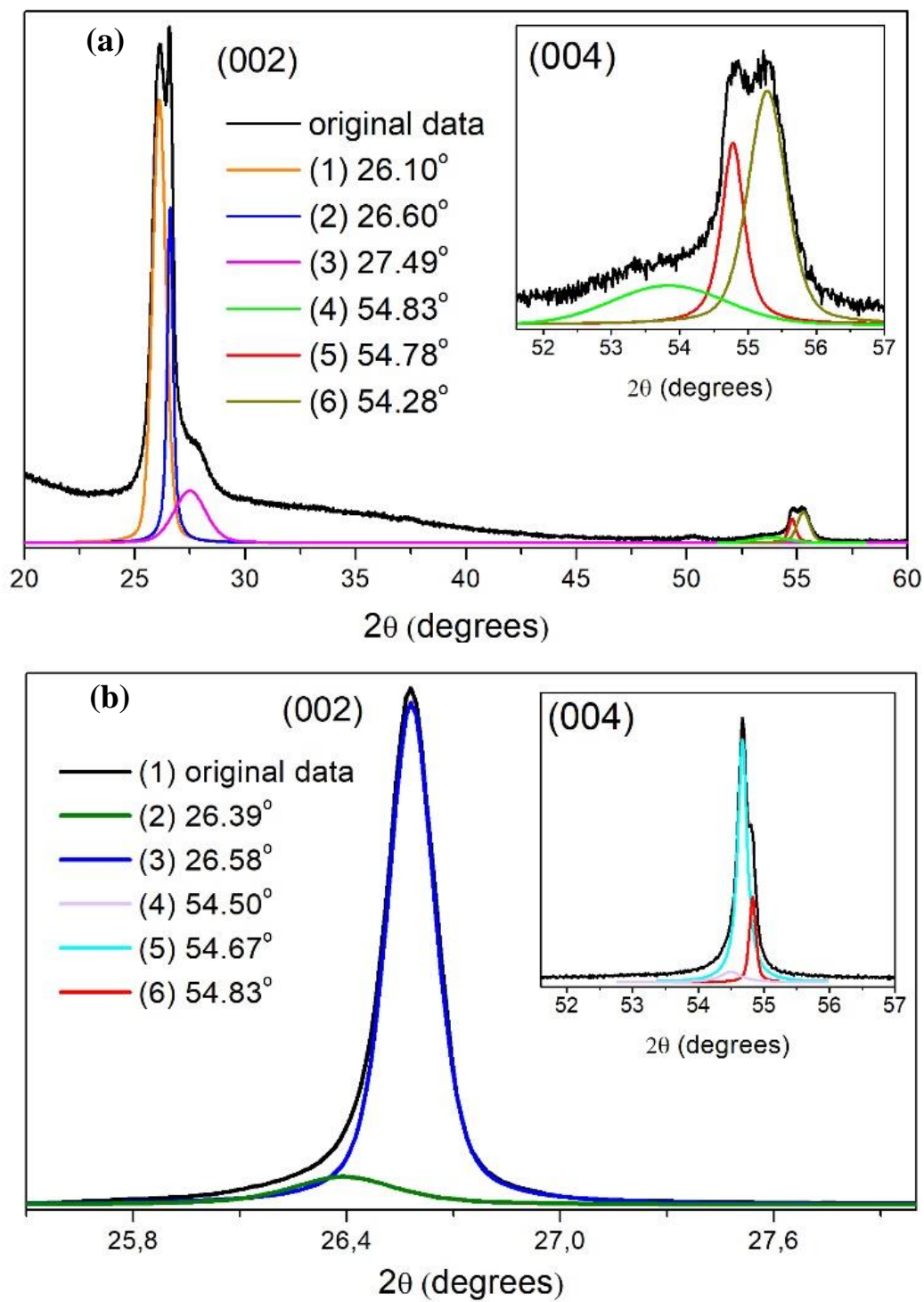

**Fig. S1.** Deconvolution examples of diffractograms obtained from GNPs powder after sonication (a) and after thermal treatment at 2950 °C (b). Legends show central positions of peaks.

**Table S1.** FWHM of (002) and (004) XRD peaks (in 2θ degrees) obtained from fitting curves after deconvolution processing.

| Sample | (002) | | | (004) | | |
|---|---|---|---|---|---|---|
| | 2H-graphite | 3R-graphite | t-graphite | 2H-graphite | 3R-graphite | t-graphite |
| Natural graphite | 0.05 | 0.10 | absent | 0.06 | 0.06 | absent |
| After sonication, not treated | 0.39 | 0.11 | 0.63 | 0.44 | 0.19 | 1.88 |
| Treated at 700 ºC | 0.16 | 0.17 | absent | 0.22 | 0.31 | absent |
| Treated at 950 ºC | 0.18 | 0.29 | absent | 0.22 | absent | absent |
| Treated at 1500 ºC | 0.18 | 0.42 | absent | 0.20 | 0.44 | absent |
| Treated at 2550 ºC | 0.18 | 0.40 | absent | 0.14 | 0.27 | absent |
| Treated at 2700ºC | 0.16 | 0.38 | absent | 0.12 | 0.15 | absent |
| Treated at 2950 ºC | 0.16 | 0.36 | absent | 0.12 | 0.17 | absent |

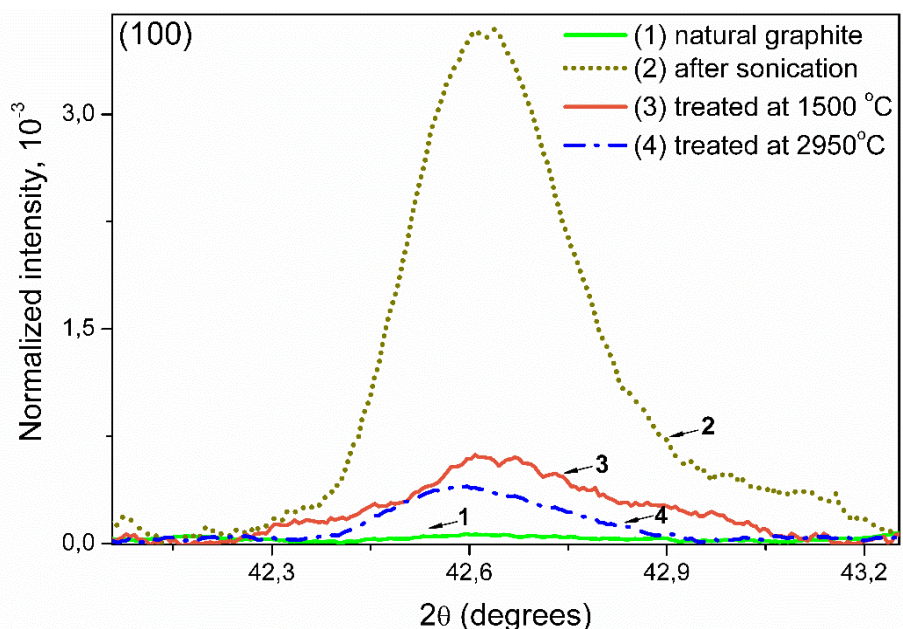

**Fig. S2.** X-ray diffraction spectra showing the (100) peak for different studied samples. Note that intensities in each spectrum are normalized by the main (002) peak intensity for the same sample.

The crystallite sizes $L_c$ and $L_a$ were calculated from (002) and (100) XRD peaks subsequently using the Scherrer equations [2]:

$$L_c = \frac{0.89 \cdot \lambda}{B_{002} \cdot \cos\theta} \text{ and } L_a = \frac{1.84 \cdot \lambda}{B_{100} \cdot \cos\theta}, \tag{1}$$

where $\lambda$ is wavelength of X-ray source, $B$ is full width half maximum of XRD peak and $\theta$ is the Bragg angle.

**Table S2.** The crystallite sizes $L_c$ and $L_a$ determined from XRD spectra.

| Sample | $L_c$ (nm) | $L_a$ (nm) |
|---|---|---|
| Natural graphite | 158 | --- |
| After sonication, not treated | 21-26 | 62 |
| Treated at 1500 ºC | 45 | 45 |
| Treated at 2950 ºC | 51 | 64 |

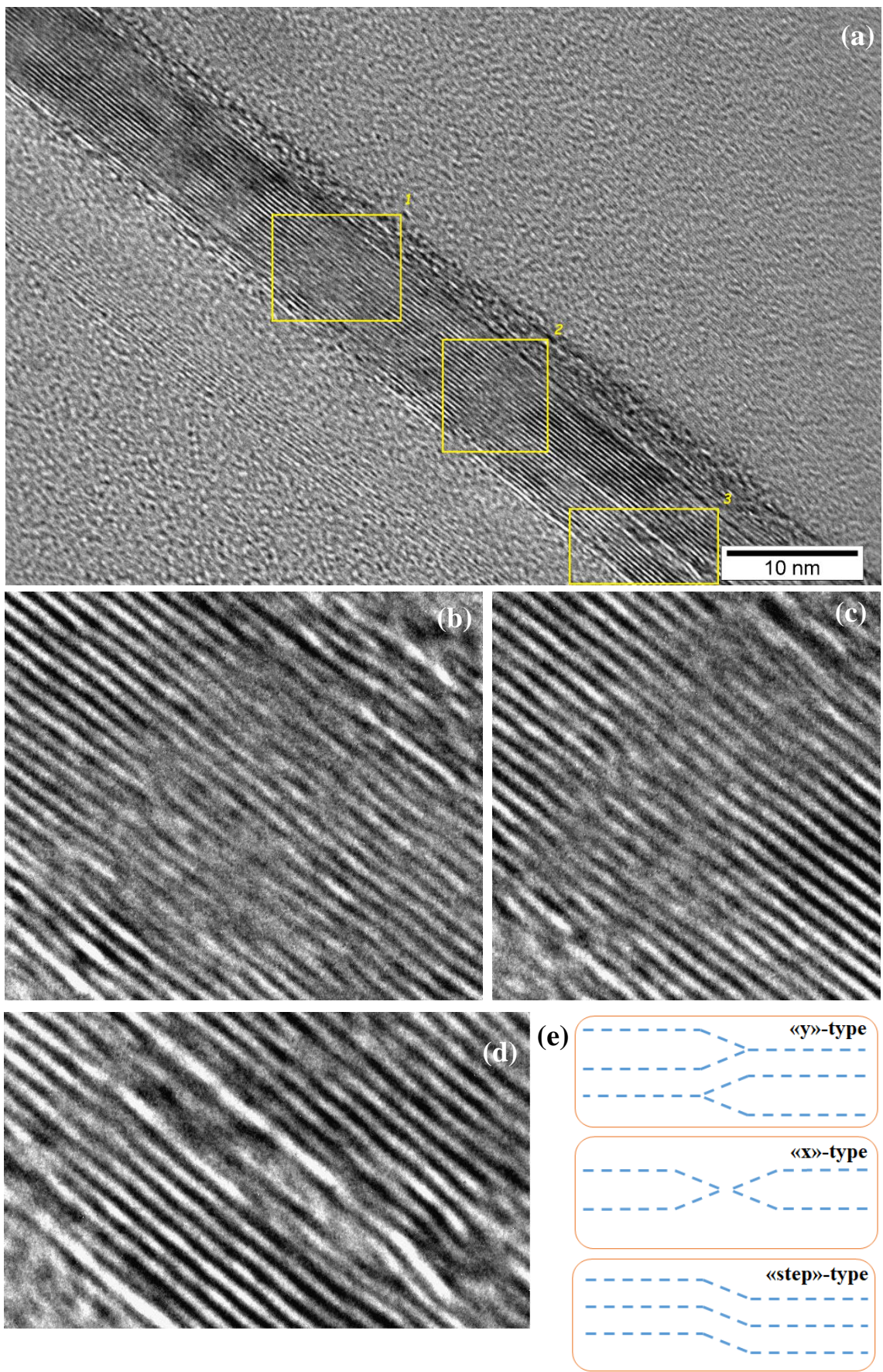

**Fig. S3.** Defects in graphite structures and their possible classification. (a) the typical HR-TEM image of GNP cross section after sonication processing. (b), (c) and (d) zoom of selected areas (1), (2) and (3) in (a) respectively. (b) «y»-type defects, (c) «x»-type and «y»-type defects and (d) «grain»-type defect. (e) schematic imaging 3 types of defects: (a) «y», (b) «x» and (c) «step»-types.

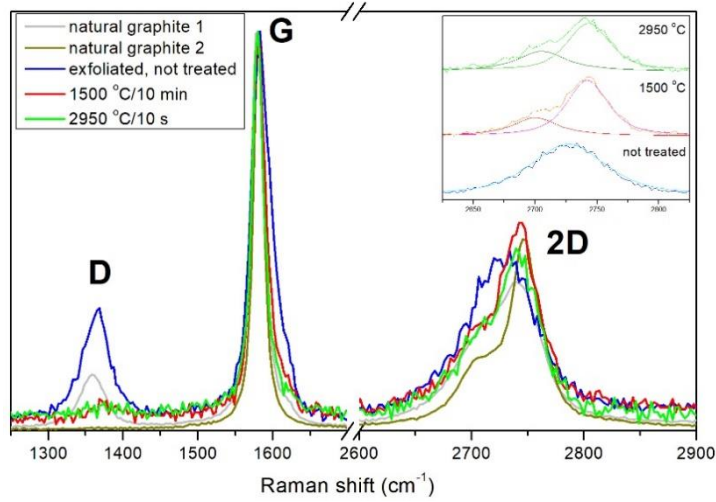

**Fig. S4.** The typical Raman spectra of GNP before and after thermal treatment. Inset shows result for 2D-band fitting.

**Table S3.** Characteristics parameters extracted from Raman spectra.

| Sample | Characteristics | | | | | | | | |
|---|---|---|---|---|---|---|---|---|---|
| | D-band position (cm$^{-1}$) | G-band position (cm$^{-1}$) | 2D-band position (cm$^{-1}$) | G-band FWHM (cm$^{-1}$) | 2D-band FWHM (cm$^{-1}$) | $I_D/I_G$ | $L_a$ (nm) | $L_D$ (nm) | $n_D$ ·10$^{10}$ (cm$^{-2}$) |
| After sonication, not treated | 1364 | 1581 | 2726 | 25 | 72 | 0.45 | 27 | 14 | 16.2 |
| Treated at 1500 °C | 1376 | 1580 | 2700 and 2742 | 18 | 43 and 39 | 0.08 | 150 | 34 | 2.9 |
| Treated at 2950 °C | 1375 | 1579 | 2705 and 2743 | 17 | 48 and 44 | 0.06 | 200 | 39 | 2.2 |
| Natural graphite 1 | 1359 | 1579 | 2707 and 2745 | 20 | 61 and 43 | 0.26 | 46 | 19 | 9,4 |
| Natural graphite 2 | absent | 1581 | 2705 and 2748 | 16 | 41 and 32 | 0 | - | - | «0» |

The in-plane crystallite size $L_a$, distance between defects $L_D$ and defect density $n_D$ were calculated using the following equations [6]:

$$L_a(nm) = 2.4 \cdot 10^{-10} \cdot \lambda_L^4 \cdot \left(\frac{I_D}{I_G}\right)^{-1}; \qquad (4)$$

$$L_D(nm) = \sqrt{1.8 \cdot 10^{-9} \cdot \lambda_L^4 \cdot \left(\frac{I_D}{I_G}\right)^{-1}}; \qquad (5)$$

$$n_D(cm^{-2}) = \frac{1.8 \cdot 10^{22}}{\lambda_L^4}\left(\frac{I_D}{I_G}\right), \qquad (6)$$

where $\lambda_L$ is the laser wavelength used in Raman spectra measurements.

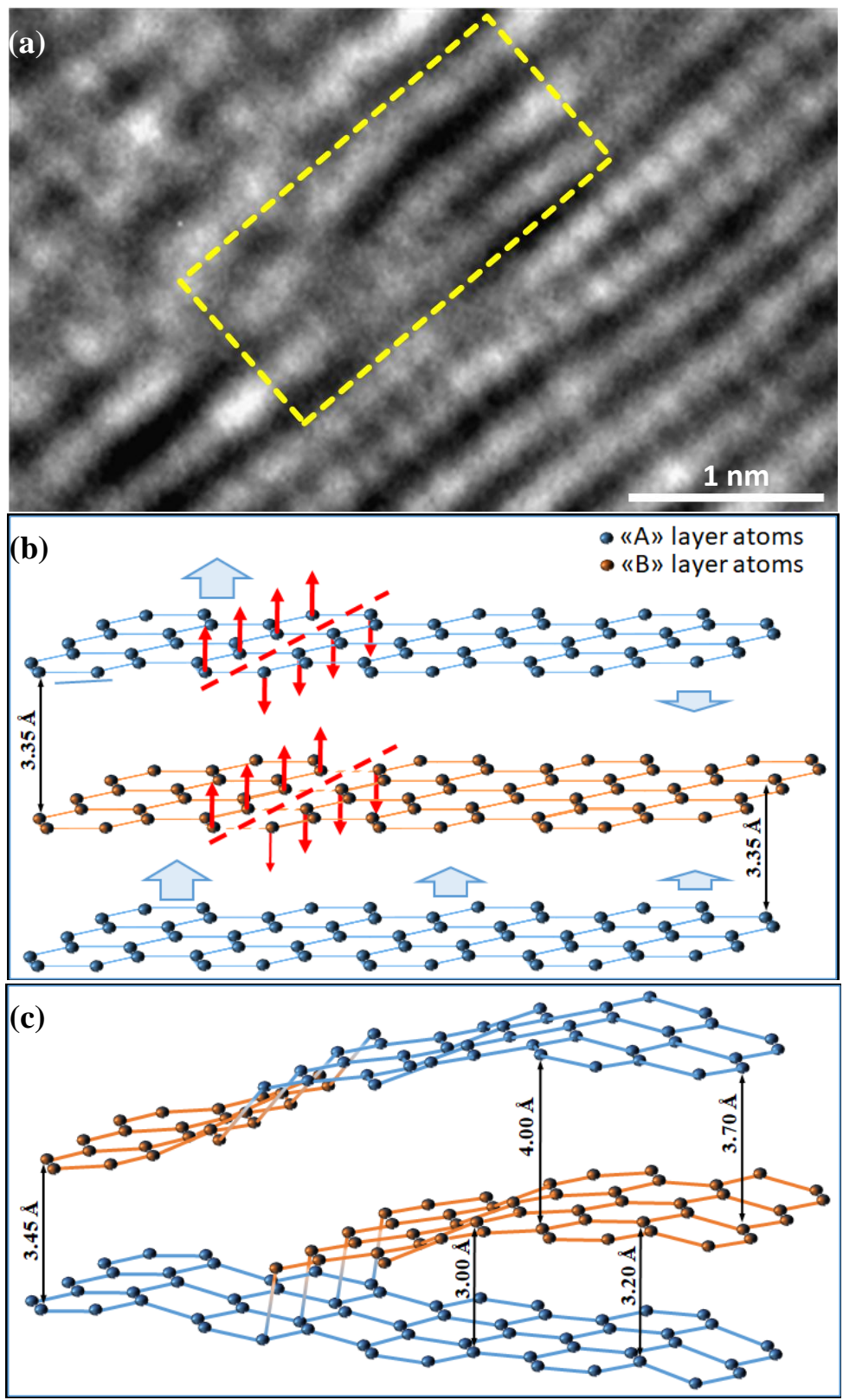

**Fig. S5.** Fragment of TEM image (a) and schematics of the buckling process after the shock-wave impact leading to breaking and re-bonding of fractured layers (b) and (c), with the positions of atoms corresponding to the area marked in (a). Dashed lines in (b) show the breaking region; large blue arrows shown direction of layer dislocation, narrow red arrows show directions of dislocation of atoms in breaking regions. Note that the distances between layers after fracture differ considerably from the initial value of 3.35 Å, both for lower and higher values.